\documentclass[twocolumn,preprintnumbers,amsmath,amssymb,aps,pra]{revtex4-1}

\usepackage[german,english]{babel}
\usepackage[utf8]{inputenc}
\usepackage{graphicx}
\usepackage{mathtools}

\usepackage{color}

\usepackage{dcolumn}% Align table columns on decimal point
\newcolumntype{d}{D{.}{.}{3}}

\newcommand*{\mat}[1]{\boldsymbol{#1}}
\newcommand*{\coord}[1]{\mathbf{#1}}
\newcommand*{\veck}{\coord{k}}
\newcommand*{\vecr}{\coord{r}}
\newcommand*{\vecs}{\coord{s}}
\newcommand*{\vect}{\coord{t}}
\newcommand*{\vecx}{\coord{x}}

\newcommand*{\binteg}[3]{\int^{\mathrlap{#3}}_{\mathrlap{#2}}\ud{#1}\,}
\newcommand*{\integ}[1]{\int\!\!\!\:\ud{#1}\:}
\newcommand*{\iinteg}[2]{\integ{#1}\!\!\!\integ{#2}}

\newcommand*{\crea}[1]{\hat{#1}^{\dagger}}
\newcommand*{\anni}[1]{\hat{#1}^{\vphantom{\dagger}}}

\newcommand{\ket}[1]{|{#1}\rangle}

\newcommand{\braket}[2]{\langle{#1}|{#2}\rangle}

\newcommand{\brakket}[3]{\langle{#1}|{#2}|{#3}\rangle}

\DeclareMathOperator{\BesselI}{\mathrm{I}}
\DeclareMathOperator{\BesselJ}{\mathrm{J}}
\DeclareMathOperator{\BesselK}{\mathrm{K}}
\DeclareMathOperator{\BesselY}{\mathrm{Y}}
\DeclareMathOperator{\Fourier}{\mathcal{F}}
\DeclareMathOperator{\Hermite}{\mathrm{H}}

\DeclareMathOperator{\Laplace}{\mathcal{L}}
\DeclareMathOperator{\Order}{\mathcal{O}}

\DeclareMathOperator{\sgn}{sgn}

\newcommand*{\abs}[1]{\lvert#1\rvert}
\newcommand*{\babs}[1]{\bigl\lvert#1\bigr\rvert}

\newcommand*{\du}{\partial}
\newcommand*{\e}{\textrm{e}}
\newcommand*{\eqspace}{\hphantom{{}={}}}

\newcommand*{\half}{\frac{1}{2}}
\newcommand*{\I}{\textrm{i}}

\newcommand*{\isDefinedAs}{\coloneqq}

\newcommand{\norm}[1]{\lVert#1\rVert}

\newcommand*{\thalf}{\tfrac{1}{2}}
\newcommand*{\ud}{\mathrm{d}}

\allowdisplaybreaks[4]

\usepackage{hyperref}

\begin{document}

\title{Natural occupation numbers: When do they vanish?}
\author{K.J.H. Giesbertz}
\affiliation{Theoretical Chemistry, Faculty of Exact Sciences, VU University, De Boelelaan 1083, 1081 HV Amsterdam, The Netherlands}
\author{R. van Leeuwen}
\affiliation{Department of Physics, Nanoscience Center, University of Jyväskylä, P.O. Box 35, 40014 Jyväskylä, Survontie 9, Jyväskylä, Finland}

\date{\today}

\begin{abstract}
The non-vanishing of the natural orbital occupation numbers of the one-particle density matrix of many-body systems has important consequences for the existence of a density matrix-potential mapping for nonlocal potentials in reduced density matrix functional theory and for the validity of the extended Koopmans' Theorem. On the basis of Weyl's theorem we give a connection between the differentiability properties of the ground state wave function and the rate at which the natural occupations approach zero when ordered as a descending series. We show, in particular, that the presence of a Coulomb cusp in the wave function leads, in general, to a power law decay of the natural occupations, whereas infinitely differentiable wave-functions typically have natural occupations that decay exponentially. We analyze for a number of explicit examples of two-particle systems that in case the wave function is non-analytic at its spatial diagonal (for instance, due to the presence of a Coulomb cusp) the natural orbital occupations are non-vanishing. We further derive a more general criterium for the non-vanishing of NO occupations for two-particle wave functions with a certain separability structure. On the basis of this criterium we show that for a two-particle system of harmonically confined electrons with a Coulombic interaction (the so-called Hookium) the natural orbital occupations never vanish.
\end{abstract}

\maketitle

%%%%%%%%%%%%%%%%%%%%%%%%%%%%%%%%%%%%%%%%
\section{Introduction}
%%%%%%%%%%%%%%%%%%%%%%%%%%%%%%%%%%%%%%%%
The fractional occupation numbers $n_j$ of the correlated one-body reduced density matrix (1RDM) have intrigued many scientists in the past decades.
They are defined by the eigenvalue equation
\begin{align}\label{eq:eigenval}
\integ{\vecx'} \gamma (\vecx,\vecx') \phi_j (\vecx') = n_j \,\phi_j (\vecx)
\end{align}
where the 1RDM itself is defined in terms of the usual creation and annihilation field operators as
\begin{align*}
\gamma(\vecx,\vecx') \isDefinedAs \brakket{\Psi}{\crea{\psi}(\vecx')\anni{\psi}(\vecx)}{\Psi}
\end{align*}
for a state $\ket{ \Psi }$ where $\vecx \isDefinedAs \vecr\sigma$ is space-spin coordinate. The one-particle orbitals $\phi_j (\vecx)$ in Eq.~\eqref{eq:eigenval} are denoted as the natural orbitals (NO)
whereas the eigenvalues $n_j$ are called the NO occupation numbers. As an integral kernel the 1RDM is a bounded linear Hermitian operator
with an infinite but countable eigenvalue spectrum and the set of all NOs form a basis in the set of quadratically integrable functions.
If the state $\ket{ \Psi }$ is fermionic it is not difficult to prove that $0 \leq n_k \leq 1$~\cite{Lowdin1955a}. In the following we will restrict ourselves 
to electronic systems such that this property holds.
The fact that the occupation numbers can also have non-integer values between zero and one
is one of the most distinct features of interacting systems compared to non-interacting systems which can only have integer occupation numbers typically. Therefore the occupation numbers reflect strongly the electronic correlations present in the system under consideration. A system is considered weakly correlating when the occupation numbers differ only slightly from zero or one, in which case the full many-electron wavefunction can well be approximated by a single Slater determinant (non-interacting wavefunction). If a system is strongly correlated, the occupation numbers deviate strongly from integer values and multiple determinants are required to obtain a sufficiently accurate approximation to the many-body wavefunction which captures the physics of the system. The ability of the 1RDM occupation numbers to signal strong correlation has encouraged people to develop 1RDM functional theory as an alternative to traditional density function theory (DFT) to handle strongly correlated systems such as dissociating molecules~\cite{GritsenkoPernalBaerends2005, RohrPernalGritsenko2008, PirisMatxainLopez2010a}, Mott insulators~\cite{SharmaDewhurst2008} and quantum Hall systems~\cite{ToloHarju2010}, for which the current approximate density functionals fail miserably.

%zero occupation numbers, interesting for 1) CI
The sum of the occupation numbers equals the number of electrons in the system. Therefore, if we order the occupation numbers, $n_k$, from the highest to the lowest one, their values need to decay to zero sufficiently fast for $k \to \infty$, i.e.
\begin{align*}
\lim_{k \rightarrow \infty} n_k = 0 ,
\end{align*}
or even become zero after some point $k_{\text{max}}$. The question whether they actually do become zero or only approach zero for $k \to \infty$ is not only an academic question, but is also of practical interest for methods that try to build an accurate approximation to the wavefunction by making an expansion in terms of Slater determinants, e.g.\ configuration interactions (CI). This question has recently been addressed for the dissociating hydrogen molecule~\cite{ShengMentelGritsenko2013}. One would expect that an optimal set of orbitals exists which leads to the fastest convergence of the expansion of the wavefunction in terms of Slater determinants~\cite{Lowdin1955a}. One can prove that if all determinants are taken into account (full CI), that the highest occupied NOs are the orbitals which give the fastest convergence towards the exact one in the $L^2$-norm~\cite{PhD-Giesbertz2010}. The NOs become even more interesting if the occupation numbers become all zero for $k$ sufficiently large, since this would imply that only a finite set of NOs would already be sufficient to expand the full many-electron wavefunction.

% 2) 1RDM functional theory
The question if zero occupation numbers exist in Coulomb systems is maybe even more important for 1RDM functional theory. 
Basic theorems in 1RDM functional theory~\cite{Gilbert1975}  follow similar arguments as the famous Hohenberg--Kohn theorem~\cite{HohenbergKohn1964}
of density functional which establishes a one-to-one correspondence between densities, potentials and non-degenerate ground states.
The main difference between 1RDM functional theory and density functional theory is that the natural conjugate variable to the 1RDM is
a non-local external potential of the form
\begin{align}\label{eq:nonloc}
\hat{V} = \iinteg{\vecx}{\vecx'} v(\vecx,\vecx') \, \crea{\psi}(\vecx ) \anni{\psi}(\vecx' )
\end{align}
rather than the local potential of density-functional theory. It therefore immediately follows that the energy contribution of the nonlocal external
field to the total energy is given by
\begin{align*}
V = \iinteg{\vecx}{\vecx'}v(\vecx,\vecx')\gamma(\vecx',\vecx),
\end{align*}
With this expression the Hohenberg--Kohn proof can be followed exactly as in density-functional theory and Gilbert~\cite{Gilbert1975} in fact 
did this to establish that there is a one-to-one correspondence between non-degenerate ground states $\ket{ \Psi }$ and their corresponding 1RDM $\gamma$. This is already sufficient to establish 1RDM functional theory, since the ground state energy can be written as a functional of the 1RDM $E[\gamma]$.
In density-functional theory one can further prove that two different (up to a gauge) potentials can not have the same non-degenerate ground state.
The analogous proof fails in 1RDM theory since there can exist nonlocal potentials $\hat{V}$ with the property that
\begin{align}\label{eq:Vannihilate}
\hat{V} \ket{ \Psi } = 0 
\end{align}
for a given ground state $\ket{ \Psi }$ of some Hamiltonian $\hat{H}$. Such a potential can therefore always be added to this Hamiltonian without affecting the ground state (it could be that $\ket{ \Psi }$ is now an excited state but by multiplying $\hat{V}$ by a small enough number we can ensure that $\ket{ \Psi }$ is still the ground state). Let us now see how Eq.~\eqref{eq:Vannihilate} can come about. Let us first define the annihilation operator
\begin{align*}
\anni{a}_s = \integ{\vecx} \anni{\psi} (\vecx ) \phi_s^* (\vecx)
\end{align*}
which annihilates the NO $\phi_s$ from any many-body quantum state. Suppose now that $\anni{a}_s \ket{ \Psi }=0$ for some of the labels $s$, which means that the orbital $\phi_s$ does not appear in any Slater determinant of a CI expansion of $\ket{ \Psi }$. This implies that
\begin{align*}
n_s = \brakket{ \Psi }{ \crea{a}_s \anni{a}_s }{ \Psi } =0
\end{align*}
such that the corresponding NO occupation number vanishes. We can then construct the following one-body potential
\begin{align*}
\hat{V} = \sum_{\mathclap{r,s \in \{i : n_i = 0\}}}  v_{rs} \, \hat{a}_r^\dagger \hat{a}_s,
\end{align*}
where $v_{rs}=v_{sr}^*$ is an arbitrary Hermitian matrix and where we sum only over the labels for which $n_r = n_s = 0$ for the state $\ket{ \Psi }$. It is clear that this potential exactly has the property $\hat{V} \ket{ \Psi }=0$. In real space this corresponds to a nonlocal spatial potential of the form of Eq.~\eqref{eq:nonloc} where
\begin{align*}
v(\vecx,\vecx') = \sum_{\mathclap{r,s \in \{i : n_i = 0\}}}  v_{rs} \, \phi_r^* (\vecx) \phi_s (\vecx').
\end{align*}
We therefore see that Eq.~\eqref{eq:Vannihilate} can be satisfied whenever the state $\ket{ \Psi }$ has vanishing NO occupations. The non-vanishing of the NO occupation numbers for electronic ground states is therefore a necessary condition for the existence of a one-to-one mapping between nonlocal potentials and 1RDMs. To the best of our knowledge the answer to the question whether the necessary condition is also a sufficient one is unknown. The one-to-one mapping between non-local potentials and 1RDMs would be relevant for the foundations of linear response 1RDM functional theory and also its time-dependent extension would greatly benefit from the resulting simplifications.

% 3) extended Koopmans
Other consequences of vanishing occupation numbers arise in the extended Koopmans' theorem~\cite{SmithDay1975, DaySmithMorrison1975, MorrellParrLevy1975, KatrielDavidson1980}. The extended Koopmans' theorem is an extension to arbitrary wavefunctions of the well known theorem by Koopmans that the occupied Hartree--Fock orbital energies provide approximations to the ionization energies~\cite{Koopmans1934}. If the exact wavefunction is used in the extended Koopmans' procedure, even the exact ionization energies should result, provided the set of partially occupied NOs is complete, i.e.\ none of the occupation numbers vanishes. A less restrictive condition has been derived by Pernal and Cioslowski~\cite{PernalCioslowski2001}, though in practice it simply implies that none of occupation numbers should vanish. For systems with Coulombic
interactions the extended Koopmans' theorem is found to hold to very high numerical accuracy~\cite{SundholmOlsen1993, OlsenSundholm1998} although this does not prove its validity.

%non-analytic behavior cusp condition
We have therefore seen that the possible vanishing of NO occupation numbers has important consequences for CI expansions as well as for the validity
of fundamental theorems in many-body theory. This then immediately raises the question in which cases the NO occupation numbers vanish. If none of the
occupation numbers vanishes, then every NO is needed in an expansion of the ground state wave function. One general observation that one can make is that
infinite expansions are typically required when expanding non-smooth functions in terms of smooth ones.
In the case of electronic ground states the Coulomb interaction requires the wavefunction to have a cusp at the positions where the electrons come together of the form
\begin{align}\label{eq:cuspCondition}
\Psi(r_{12} \to 0) = \Psi(r_{12} = 0)\left(1 + \half r_{12} + \dotsb\right),
\end{align}
where $r_{12} \isDefinedAs \abs{\vecr_1 - \vecr_2}$. This cusp gives an infinite kinetic energy which exactly compensates the infinity from the Coulomb interaction between the electrons~\cite{Lowdin1954, Kato1957, Bingel1963, PackBrown1966}. Due to this non-analytic behavior of the wavefunction, a full expansion of the wavefunction in one-electron functions requires in general all functions to be present. Hence, one may expect in the particular case of an expansion in NOs, none of the NOs should have an occupation number equal to zero, since that would imply that the NO is not required in the expansion.

%made exact for HEG (Kimball)
Although this argument sounds very reasonable, it is certainly not a proof that zero occupations do not occur in Coulomb systems. Though infinitely many occupation numbers are required to be non-zero, it might be that some of them are still zero in some special situations. In the case of the homogeneous electron gas (HEG), however, this argument can be turned into a proof. Since the NOs of the HEG are simply plane waves, the occupation numbers are then given by the momentum distribution, $n(k)$. Kimball has shown that the momentum distribution is required to decay as $1/k^8$ due to the inter-electronic cusp condition~\cite{Kimball1975}, so the occupation numbers never become exactly zero.

%intro 2-elec systems
In the case of the HEG we were in the fortunate situation that the NOs are plane waves and, so that their occupation numbers are simply given by the momentum distribution. For general systems we are not in such a convenient position, because a straightforward expansion in a finite basis set effectively smoothens the electron-electron cusp~\eqref{eq:cuspCondition} and the argument does not apply anymore. For two-electron systems we are in a more fortunate situation, however, since for singlet two-electron systems there is a strong connection between the NOs and the wavefunction. The spatial part of the singlet two-electron wavefunction is symmetric and can therefore be diagonalized
\begin{align}\label{eq:PsiSpectral}
\Psi(\vecr_1,\vecr_2) = \sum_kc_k\phi_k(\vecr_1)\phi_k(\vecr_2).
\end{align}
By calculating the corresponding spin-integrated 1RDM, one readily finds that the eigenfunctions are NOs and that the coefficients are related to the occupation numbers as $n_k = c_k^2$. Though we are not in such a good position as the for the HEG, this connection is quite useful, since it allows us to connect the behavior of the occupation numbers directly to the analytic properties of the wavefunction, instead of going via the 1RDM in which much of the analytic properties are integrated out. Therefore, we will focus our attention in this paper mainly to singlet two-electron systems to demonstrate how the form of the interaction determines the analytic properties of the wavefunction, which in turn dictates the asymptotic decay of the occupation numbers.

%%%%%%%%%%%%%%%%%%%%%%%%%%%%%%%%%%%%%%%%
\section{Explicit examples}
%%%%%%%%%%%%%%%%%%%%%%%%%%%%%%%%%%%%%%%%
Before we present a general treatment for simple explicitly correlated wavefunctions, we first consider some specific examples for which we can solve the NOs and coefficients explicitly or at least prove that none of the NOs have a vanishing coefficient (occupation number).

%Hylleraas 1D
\subsection{A simple 1D Hylleraas wavefunction}
\label{sec:Hylleraas1D}
Let us first consider the simplest wavefunction with a cusp in one dimension (1D)
\begin{align*}
\Psi(x_1,x_2) = K\alpha(x_1)\alpha(x_2)(1 + \eta\abs{x_1 - x_2}),
\end{align*}
where $K$ is a normalization constant and $\alpha(x)$ is an arbitrary orbital apart from the fact that it is positive, $\alpha(x) > 0$. To calculate the NOs and the coefficients in the spectral expansion of the wavefunction~\eqref{eq:PsiSpectral}, we need to solve the following eigenvalue equation
\begin{multline*}
K \integ{x_2}\alpha(x_1)\alpha(x_2)(1 + \eta\abs{x_1 - x_2})\phi_k(x_2) \\
= c_k\,\phi_k(x_1).
\end{multline*}
Introducing the following function $\varphi_k(x) \isDefinedAs \phi_k(x)/\alpha(x)$
the eigenvalue equation can be written as
\begin{align*}
K \integ{x_2}(1 + \eta\abs{x_1 - x_2})\alpha^2(x_2)\varphi_k(x_2) = c_k\,\varphi_k(x_1).
\end{align*}
Now differentiating this equation twice with respect to $x_1$, we obtain the following differential equation for $\varphi_k(x)$
\begin{align*}
c_k \varphi_k''(x) = 2\eta K \alpha^2(x) \varphi_k(x),
\end{align*}
where we used that $\abs{x}'' = 2\delta(x)$. From this equation we see that $c_k \neq 0$ if $\eta \neq 0$, since otherwise we would have $\varphi_k = 0$ which is not an eigenfunction. Hence, for $\eta \neq 0$ we can divide by $c_k$ and write the equation as
\begin{align*}
\varphi_k''(x) = \lambda_k \alpha^2(x) \varphi_k(x),
\end{align*}
where
\begin{align}\label{eq:lambdaDef}
\lambda_k \isDefinedAs 2\eta K/c_k.
\end{align}
In the case that a simple Slater function is used for the orbital, $\alpha(x) = \e^{-Z\abs{x}}$, the differential equation can be cast into a Bessel's differential equation. The full construction of all the NOs and their coefficients is rather technical and has been deferred to Appendix~\ref{ap:modelAtom}.
The result of the calculation is that none of the occupation numbers is zero and that the occupation numbers behave asymptotically as
\begin{align*}
n_k \sim \frac{C}{k^2} \quad (k \rightarrow \infty)
\end{align*}
where $C$ is a constant.
We will see later that such a power law behavior is typical for wave functions that are at most a finite number of times differentiable
(in our case zero times). 

%nabla^4 proof
\subsection{A simple 3D Hylleraas wavefunction}
Let us now try to extent the approach of the previous example to a three dimensional case, so that we have a wavefunction in three dimensions (3D) that satisfies the cusp condition~\eqref{eq:cuspCondition} of the form
\begin{align*}
\Psi(\vecr_1,\vecr_2) = K\,\alpha(\vecr_1)\alpha(\vecr_2)(1+ \eta\, r_{12})
\end{align*}
where $r_{12}=|\vecr_1 - \vecr_2|$.
Since the NOs are eigenfunctions of the wavefunction~\eqref{eq:PsiSpectral}, they satisfy the eigenvalue equation
\begin{align}\label{eq:PsiEgen}
\integ{\vecr'}\Psi(\vecr,\vecr')\phi_k(\vecr') = c_k\,\phi_k(\vecr).
\end{align}
Now introducing similar function as in the 1D case, $\varphi_k(\vecr) \isDefinedAs \phi_k(\vecr)/\alpha(\vecr)$, the eigenvalue equation can be written as
\begin{align*}
K\integ{\vecr'}\bigl(1+ \eta\abs{\vecr-\vecr'}\bigr)\alpha^2(\vecr')\varphi_k(\vecr') = c_k\,\varphi_k(\vecr).
\end{align*}
Unfortunately, just taking the Laplacian does not work, since $\nabla_{\vecr}^2\abs{\vecr-\vecr'} = 2/\abs{\vecr-\vecr'}$. Taking the Laplacian a second time, however, we obtain the sought after delta function and our equation becomes
\begin{align*}
-8\pi\eta K\,\alpha^2(\vecr)\varphi_k(\vecr) = c_k\nabla^2 \nabla^2 \varphi_k(\vecr).
\end{align*}
We can now use the same argument as in the 1D case. Since the orbital $\alpha(\vecr)$ will in general not vanish on some open set, $c_k = 0$ would imply that $\phi_k(\vecr) = 0$ for unoccupied NOs. Since such orbitals are not normalizable, our simple explicitly correlated wavefunction does not have any NOs with occupation numbers equal to zero. Since the differential equation for the functions $\varphi_k(\vecr)$ is now fourth order and additionally in 3D, it becomes quite hard to obtain the NOs explicitly from this equation. In any case, it is exactly the cusp behavior that allows us to conclude that none of the
NO occupation numbers vanish.

%Also harmonium ni ≠ 0 -> more general?
\subsection{Double Harmonium}
\label{sec:harmInt}
Although we have found that the exact treatment of the cusp in the simple Hylleraas wavefunction prevented the NOs to have a zero occupation number, one can wonder if the electron-electron cusp is actually essential to have only non-zero occupation numbers. This is actually not the case as one can show explicitly for a system with harmonic interactions in one dimension. The Hamiltonian for such a system can be written as
\begin{align*}
\hat{H} = -\half\frac{\du^2}{\du x_1^2} - \half\frac{\du^2}{\du x_2^2} + 
\frac{\omega^2}{2}(x_1^2 + x_2^2) + \lambda(x_1-x_2)^2,
\end{align*}
where $\lambda \geq 0$. The coordinates can be decoupled by making a transformation to center-of-mass coordinates, which gives the following expression for the Hamiltonian
\begin{align}\label{eq:HhamRel}
\hat{H} = -\half\frac{\du^2}{\du s^2} + \half\omega^2s^2  - 
\half\frac{\du^2}{\du t^2}+ \half\tilde{\omega}^2 t^2,
\end{align}
where $s \isDefinedAs (x_1 + x_2)/\sqrt{2}$, $t \isDefinedAs (x_1 - x_2)/\sqrt{2}$ and $\tilde{\omega}^2 \isDefinedAs \omega^2 + 4\lambda$. Since this is the Hamiltonian for two independent harmonic oscillators, we can immediately write spatial part of the singlet ground state wavefunction as
\begin{align*}
\Psi(x_1,x_2)
&= \sqrt[4]{\frac{\omega\,\tilde{\omega}}{\pi^2}}\e^{-\half(\omega s^2 + \tilde{\omega}t^2)} \notag \\
&= \sqrt[4]{\frac{\omega\,\tilde{\omega}}{\pi^2}}\e^{-\half\omega x_1^2}\e^{-\half\omega x_2^2}
\;\e^{-\half(\tilde{\omega}-\omega)t^2}.
\end{align*}
The structure of this solution is sufficiently simple to determine the NOs and the corresponding wavefunction coefficients explicitly. Due to the purely harmonic nature of our system, one might suspect that the NOs also have the form of harmonic oscillator solutions. Indeed, one can actually calculate the NOs to be (see Appendix~\ref{ap:harmNOs})
\begin{align}\label{eq:harmNOs}
\phi_k(x) = a_k\Hermite_k\bigl(\sqrt[4]{\omega\tilde{\omega}}\,x\bigr)
\e^{-\half\sqrt{\omega\tilde{\omega}}x^2},
\end{align}
where $\Hermite_k(x)$ are the Hermite polynomials and $a_k$ normalization constants satisfying
\begin{align*}
a_k^2 = \frac{\sqrt[4]{\omega\tilde{\omega}}}{2^kk!\sqrt{\pi}}.
\end{align*}
The corresponding wavefunction coefficients are
\begin{align}\label{eq:harmCoefs}
c_k = \frac{2\sqrt[4]{\omega\tilde{\omega}}}{\sqrt{\omega}+\sqrt{\tilde{\omega}}}\, (-1)^k
\biggl(\frac{\sqrt{\omega}-\sqrt{\tilde{\omega}}}{\sqrt{\omega}+\sqrt{\tilde{\omega}}}\biggr)^k.
\end{align}
We find that the wavefunction coefficients are alternating and only vanish in the limit $k \to \infty$. Only when $\tilde{\omega} = \omega$ (no interactions), all the coefficients become zero, except for the first one $c_0=1$. An important conclusion is now that instead of a power law behavior we now have 
an exponential decay of the form
\begin{align*}
n_k = C \, a^k 
\end{align*}
with $C$ and $a$ constants. We will show below that such a behavior is typical in the limit $k \rightarrow \infty$ 
for infinitely differentiable wave functions.
Our examples seem to indicate that there is a connection between the differentiability properties of the wave functions and 
the asymptotic behavior of the NO occupations. In the next Section we make this connection more precise.

\section{A lower bound on the decay rate of the occupation numbers}
The eigenvalue equation for the NOs~\eqref{eq:PsiEgen} is a Fredholm integral equation, where the two-body wavefunction $\Psi$ is the kernel and $c_k$ are the eigenvalues. (Often one considers the characteristic values, $1/c_k$, instead of the eigenvalues.) In the theory of Fredholm integral equations lower bounds on the decay rate of the eigenvalues have been established on the differentiability and analyticity of the integral kernel~\cite{Weyl1911, HilleTamarkin1931, Rasmussen2001}. Although only lower bounds have been found, these bounds are a nice illustration of the strong link between the differentiability of the kernel and the decay rate of the kernel.

The first result of interest is by Hille and Tamarkin, who showed that the eigenvalues of operators with analytic kernels decay exponentially. More precisely
\begin{align*}
\lim_{k \to \infty}\abs{\lambda_k}\,R^{\frac{1}{4}k} = 0,
\end{align*}
where $\lambda_k$ are the eigenvalues and the constant $R$ is related to the size of the region where the kernel is analytic~\cite{HilleTamarkin1931, Rasmussen2001}. Indeed, the kernel of the double harmomium is analytic and the corresponding coefficients decay exponentially~\eqref{eq:harmCoefs}.

A result for finitely differentiable integral kernels by Weyl~\cite{Weyl1911} is of particular interest for wavefunctions with a cusp.
Weyl showed that the eigenvalues of a finitely differentiable kernel only need to decay polynomially. More precisely, if the partial derivatives of a symmetric kernel are continuous up to order $p$, the decay rate of its eigenvalues is bounded by
\begin{align}\label{eq:WeylDecay}
\lim_{k \to \infty}\abs{\lambda_k}\,k^{p/d + \half} = 0,
\end{align}
where $d$ is the dimension of the integration variable. A derivation of Weyl's result can be found in Appendix~\ref{ap:WeylTheorem}.

Let us compare Weyl's theorem with the results for the simple 1D Hylleraas atom (Sec.~\ref{sec:Hylleraas1D}). The first derivative of the wavefunction is already discontinuous in this case, so $p=0$. This gives us a rather modest lower bound on the decay rate of the coefficients, comparing to the actual quadratic decay-rate $\bigl(k^{-2}\bigr)$ of the coefficients (see Appendix~\ref{ap:modelAtom}). The bound by Weyl can be tightened by using more regularity properties of the discontinuous derivatives of the wavefunction~\cite{HilleTamarkin1931, Rasmussen2001}. Since the discontinuity is bound, the two-electron wavefunction can be put in $\text{Lip}_1(1,2)$~\cite{HilleTamarkin1931, Rasmussen2001} which gives the lower bound $k^{-3/2}$ on the decay-rate of the coefficients.

Though Weyl's theorem does not put a very stringent constraint on the decay of the occupation numbers, it clearly demonstrates that wavefunctions with an interelectronic cusp will typically have only a polynomial decaying occupation number spectrum, whereas the occupation numbers of a wavefunction without a cusp will decay exponentially. Since in a finite basis set the non-analyticity of the cusp can not be fully represented and thereby effectively smoothened, the calculated occupation numbers will always decay exponentially or faster, i.e.\ too fast compared to the typical polynomial decay. This is actually not surprising, since if only a finite number of orbitals is included in the calculations, all the orbitals in the complement are automatically NOs with zero occupation number. Additionally, since a finite basis set representation effectively removes the cusp from the wavefunction~\eqref{eq:cuspCondition}, the original argument that NOs with vanishing occupation numbers do not exist in many-body Coulomb systems, does not even apply anymore.

%Fourier proof
%%%%%%%%%%%%%%%%%%%%%%%%%%%%%%%%%%%%%%%%
\section{A proof by Fourier transform}
%%%%%%%%%%%%%%%%%%%%%%%%%%%%%%%%%%%%%%%%
The example with the harmonic interaction shows that the cusp actually might not be essential to prevent occupation numbers becoming zero. It seems that any correlation that requires some $r_{12}$ behavior that can not be expressed in a finite number of simple orbital products will probably imply the absence of unoccupied NOs. We can make this idea more precise in the case of a singlet two-electron system which is limited to the form of a simple orbital product times an arbitrary correlation function depending only on $\vecr_1 - \vecr_2$, i.e.\ a wavefunction of the form
\begin{align}\label{eq:f12wave}
\Psi(\vecr_1,\vecr_2) = \alpha(\vecr_1)\alpha(\vecr_2)f(\vecr_1 - \vecr_2),
\end{align}
where $\alpha(\vecr) > 0$. Considering the situation that an NO, $\phi_i(\vecr)$, has a zero occupation number (so also $c_i = 0$), the eigenvalue equation~\eqref{eq:PsiEgen} then simplifies to
\begin{align*}
\integ{\vecr'}f(\vecr-\vecr') \, \chi_i(\vecr') = 0,
\end{align*}
where $\chi_i(\vecr) \isDefinedAs \alpha(\vecr)\phi_i(\vecr)$. Since this condition has the form of a convolution product, we can deconvolute it by taking the Fourier transform
\begin{align}
\tilde{f}(\veck)\tilde{\chi}_i(\veck) = 0.
\end{align}
Provided that $\tilde{f}(\veck) \neq 0$ almost everywhere, it follows that $\tilde{\chi}_i(\veck) = 0$. Since $\alpha(\vecr) \neq 0$ almost everywhere, this implies that the NO $\phi_i(\vecr) = 0$. Because this is not a normalizable function, we can conclude that no eigenvalues $c_i = 0$ exist, if $\tilde{f}(\veck) \neq 0$ almost everywhere.

The converse also holds. If $\tilde{f}(\veck) = 0$ on some finite interval, an NO with zero occupancy exists by construction. To construct this NO, define a $\tilde{\chi}_k(\veck)$ which is non-zero on this interval where $\tilde{f}(\veck)$ vanishes. By Fourier transforming back, we can construct the corresponding NO ($\phi_k(\vecr) = \chi_k(\vecr)/\alpha(\vecr)$). Since this will be a function in the null-space of the (linear) 1RDM-operator, it can not be expressed linear combination of NOs with finite occupancy. Let us show this more explicitly. Suppose that such a linear combination exists, then we write the new NO as
\begin{align*}
\phi_k(\vecr) = \smashoperator{\sum_{i\in \{i : n_i \neq 0\}}}b_i\,\phi_i(\vecr).
\end{align*}
Since the 1RDM-operator is linear, we can write its action on $\phi_k$ as
\begin{align*}
0 = \integ{\vecr'}\gamma(\vecr,\vecr')\phi_k(\vecr')
= \smashoperator[l]{\sum_{i\in \{i : n_i \neq 0\}}}\integ{\vecr'}\gamma(\vecr,\vecr')\phi_i(\vecr')b_i.
\end{align*}
If we now take the inner product of this result with $\phi_k$, we find
\begin{align*}
0 = \smashoperator{\sum_{i,j \in \{i : n_i \neq 0\}}}b_i^*b_jn_i\braket{\phi_j}{\phi_i}
= \smashoperator{\sum_{i \in \{i : n_i \neq 0\}}}n_i\abs{b_i}^2.
\end{align*}
Hence, we have a contradiction, so the constructed NO with zero occupation number can not be expressed as a linear combination of the NOs with a finite occupancy.

%works for Hylleraas and harmonium
Now let us check if our proof indeed recovers the result that our previous examples do not have vanishing occupation numbers. In the case of the harmonium the correlation function is simply a Gaussian. Since the Fourier transform of a Gaussian is simply again a Gaussian, the Fourier transform is non-zero every where. Hence, no unoccupied NOs should exist, which is in agreement with our explicit construction. The situation is more complicate in the case of the simple Hylleraas wavefunction. The correlation function is not an $L^2$ function anymore, so we can expect distributions to appear in its Fourier transform~\footnote{The corresponding test function space can probably be constructed from the NOs as $\chi_k(\vecr) = \alpha(\vecr)\phi_k(\vecr)$.}. Note that a divergent correlation function is allowed, provided the divergence of the correlation function is compensated by a stronger decay of the orbital $\alpha(\vecr)$, to make the wavefunction normalizable. The Fourier transform in 1D becomes
\begin{align}\label{eq:Hylleraas1Dfourier}
\Fourier[1 + \eta\abs{x}](k) = 2\pi\delta(k) - \frac{2\eta}{k^2}
\end{align}
and in 3D we find
\begin{align}\label{eq:Hylleraas3Dfourier}
\Fourier[1 + \eta r](k) = -\frac{4\pi}{k}\left(\frac{2\eta}{k^3} + \pi\delta'(k)\right).
\end{align}
The calculation of these Fourier transforms has been worked out in more detail in Appendix~\ref{ap:Fouriers}. Since we find in both cases that the Fourier transform is non-zero everywhere, we recover the result that no unoccupied NOs exists as we have found before. Now let us apply the theorem to some other systems.

\subsection{Inverse harmonic interaction}
\label{sec:invHarmInter}
First we will consider a system with inverse harmonic interactions which has been considered before by Morrison et al.~\cite{MorrisonZhouParr1993}. The inverse harmonic interaction is also sometimes referred to as the Calogero interaction~\cite{Calogero1969a, Calogero1969b, Calogero1971}. The full Hamiltonian we will consider here is given as
\begin{align*}
\hat{H} = -\half\nabla_{\vecr_1}^2 - \half\nabla_{\vecr_2}^2 +
\half\omega^2(r_1^2 + r_2^2) + \frac{\lambda}{r_{12}^2}.
\end{align*}
The eigenstates of this Hamiltonian can be solved exactly by making a transformation to the centre-of-mass coordinates, which decouples the coordinates. The Hamiltonian for the centre-of-mass coordinate is simply the Hamiltonian of a harmonic oscillator, so is readily solved. The Hamiltonian for the relative coordinate is more involved, but can still be solved in terms of confluent hypergeometric functions~\cite{MorrisonZhouParr1993, LandauLifshitz1977p127-128}. Fortunately, the ground state reduces to the following particularly simple form
\begin{align*}
\Psi(\vecr_1,\vecr_2) = \sqrt{\frac{\omega^{3+\alpha}}{2^{1+\alpha}\pi^{5/2}\Gamma\bigl(\frac{3}{2}+\alpha\bigr)}}\e^{-\half\omega(r_1^2+r_2^2)}r_{12}^{\alpha},
\end{align*}
where $\Gamma(z)$ is the gamma function and $\alpha = \bigl(\sqrt{1+4\lambda}-1\bigr)/2$~\footnote{The factor $1/2$ in the exponent seems to be missing in Ref.~\cite{MorrisonZhouParr1993}. This error propagates throughout the article.}. Since the ground state is simply the product of an orbital, $\alpha(\vecr_1)\alpha(\vecr_2)$, times a correlation function, $f(\vecr_{12}) = r_{12}^{\alpha}$, our theorem can be applied to this case. For the Fourier transform of the correlation function we find (see Appendix~\ref{ap:Fouriers} for details)
\begin{multline}\label{eq:rAlphaFourier}
\Fourier[r^{\alpha}](k) \\
= -\frac{4\pi}{k} \times \begin{cases}
\pi(-1)^{\alpha/2}\delta^{(1+\alpha)}(k)				&\text{even $\alpha$} \\
\dfrac{\Gamma(2 + \alpha)\sin(\pi\alpha/2)}{k^{2+\alpha}}	&\text{otherwise},
\end{cases}
\end{multline}
where $\delta^{(n)}(k)$ denotes the $n^{\text{th}}$ order derivative of the delta-function. Therefore, we find that zero occupation numbers can only exist for even $\alpha$, which is in agreement with the findings of Morisson et al.~\cite{MorrisonZhouParr1993}. Actually, it is not surprising that there are only a finite number of unoccupied NOs for even $\alpha$, since in that case the correlation function becomes exactly separable in $r_1$ and $r_2$, so the wavefunction can be represented by a finite number of orbitals. For example in the simplest non-trivial case of $\alpha = 2$ the wavefunction can be written as
\begin{multline*}
\Psi(\vecr_1,\vecr_2) = \frac{\omega}{\sqrt{15}}\biggl(\frac{\omega}{\pi}\biggr)^{\mathrlap{3/2}}\,
\bigl[\chi_{1}(\vecr_1)\chi_{r^2}(\vecr_2) + \chi_{r^2}(\vecr_1)\chi_{1}(\vecr_2) \\
{} - 2\bigl(\chi_{x}(\vecr_1)\chi_{x}(\vecr_2) + \chi_{y}(\vecr_1)\chi_{y}(\vecr_2) + \chi_{z}(\vecr_1)\chi_{z}(\vecr_2)\bigr)\bigr],
\end{multline*}
where we used $\chi_{f}(\vecr) \isDefinedAs f(\vecr) \e^{-\half \omega r^2}$ as a compact notation for the various one-particle functions. Since only five orbitals are required to represent this wavefunction, only five NOs with non-zero occupation number exist. We see that the contribution from the $p$ orbitals to the wavefunction is already diagonal, so to obtain the NOs, we only need to normalize them
\begin{align*}
\phi_x(\vecr) = \sqrt{2 \omega\left(\frac{\omega}{\pi}\right)^{3/2}}\;x\,\e^{-\half\omega r^2}
\end{align*}
and we have similar expressions for $\phi_y(\vecr)$ and $\phi_z(\vecr)$ of course~\footnote{There is a subtle difference between our result and the one in Ref.~\cite{MorrisonZhouParr1993}. Since the spectral representation of the wavefunction~\eqref{eq:PsiSpectral} does not have a complex conjugation, their NOs with spherical harmonics does not diagonalize the wavefunction, though it does diagonalize the 1RDM. Using the cartesian version we diagonalize both the wavefunction and the 1RDM.}. Their coefficient in the spectral expansion~\eqref{eq:PsiSpectral} is readily obtained as $c_x = c_y = c_z = -1/\sqrt{15}$. The contribution from the other two orbitals $\chi_{1}(\vecr)$ and $\chi_{r^2}(\vecr)$ is not diagonal, so has to be diagonalized. This is readily achieved by the following linear combinations
\begin{align*}
\phi_{\pm}(\vecr)
= \sqrt{\frac{2a}{4a \pm 3}\left(\frac{\omega}{\pi}\right)^{3/2}}\left(\frac{\omega}{2a}r^2 \pm 1\right)\e^{-\half\omega r^2},
\end{align*}
where $a \isDefinedAs \sqrt{15}/4$. The orbitals $\phi_{\pm}(\vecr)$ are constructed such that they are orthonormal, so they are the required NOs. Their corresponding expansion coefficients can be calculated to be $c_{\pm} = (4a \pm 5)/10$.

This example clearly demonstrates that the cusp is actually not essential for the absence of unoccupied NOs. The absence of zero occupation numbers is caused by correlation in the full many-body wavefunction, which can not be expanded in a finite series of one-electron functions. The cusp actually causes the exact wavefunction to have such a form, hence there will be an infinite amount of non-zero occupation numbers.

%hookium
\subsection{Hookium atom}
In this section we will apply our theorem to Hooke's atom. The interaction is now the Coulomb interaction, the interaction of interest for electrons. However, the confining potential is still harmonic to allow for the separation of variables by changing the coordinates to the centre-of-mass frame. The Hamiltonian is given as
\begin{align*}
\hat{H} = -\half\nabla_1^2 - \half\nabla_2^2 + \half\omega^2(r_1^2 + r_2^2) + \frac{\lambda}{r_{12}}.
\end{align*}
The ground state has the following form
\begin{align*}
\Psi(\vecr_1,\vecr_2) = N\e^{-\half\omega(r_1^2+r_2^2)}\frac{t\bigl(\sqrt{\omega/2}\,r_{12}\bigr)}{r_{12}},
\end{align*}
where $N$ is a normalization constant. The function $t(\rho)$ satisfies the following differential equation~\cite{KestnerSinanoglu1962, Taut1993} for the ground state
\begin{align*}
\rho t'' - 2\rho^2t'+\biggl((\tilde{\epsilon}_r-1)\rho - \frac{\lambda}{\sqrt{\omega/2}}\biggr)t = 0,
\end{align*}
where $\tilde{\epsilon}_r = 2\epsilon_r/\omega$ with $\epsilon_r$ as the contribution to the energy from the relative coordinate (the total energy is $\frac{3}{2}\omega + \epsilon_r$). Unfortunately, this differential equation does not allow for an explicit solution, though a series solution can be constructed which even has only a finite number of terms for specific ratios $\lambda^2/\omega$. More importantly, by neglecting the second order derivative and the last term, we readily find that the solution has to behave asymptotically for large $\rho$ as
\begin{align*}
t(\rho \to \infty) \sim \rho^{\frac{\epsilon_r}{\omega} - \half},
\end{align*}
so at least its Fourier transform exists as a distribution and in particular its Laplace transform
\begin{align*}
\tilde{t}(s) \isDefinedAs \Laplace[t](s) \isDefinedAs \binteg{\rho}{0}{\infty}\e^{-s\rho}t(\rho)
\end{align*}
exists. The differential equation for $t(\rho)$ can be transformed into the following differential equation for the Laplace transform
\begin{multline}\label{eq:hookLaplEq}
2s\tilde{t}''(s) + \left(s^2 + \epsilon_r + 3\right)\tilde{t}'(s) \\
{} + \biggl(2s + \frac{\lambda}{\sqrt{\omega/2}}\biggr)\tilde{t}(s) = 0,
\end{multline}
where we used that $t(0) = 0$. The Fourier transform of the correlation function can now directly be obtained from $\tilde{t}(s)$ as
\begin{align}\label{eq:FourLapl}
\Fourier[f](k)
&= \frac{2\pi \I}{k}\lim_{\sigma \to 0^+}\left(\tilde{t}(\sigma+\I k) - \tilde{t}(\sigma-\I k)\right),
\end{align}
where the limits are important to include possible poles at the origin. Since the Laplace transform is a solution of a second order differential equation, it has to be an analytic function. This analyticity is carried over to the Fourier transform except for some possible irregularities located at the origin ($k=0$). Since an analytic function can only be zero in an open set if it vanishes everywhere, the Fourier transform of the correlation function of the Hookium atom does not vanish in a finite region and hence, the NO coefficients (occupation numbers) do not vanish for the Hookium atom.

\begin{figure}[t]
  \includegraphics[width=\columnwidth]{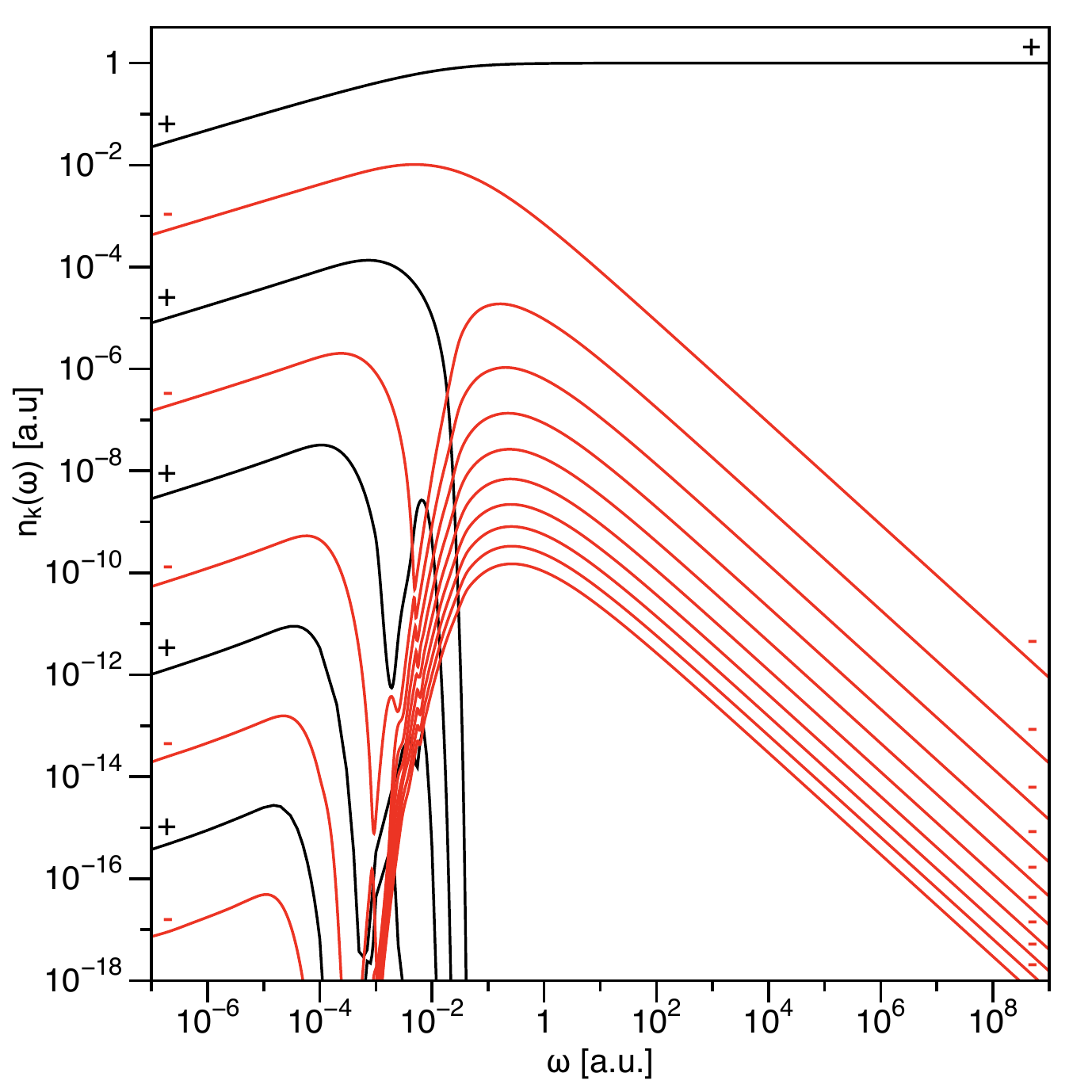}
  \caption{The NO occupations of the harmonium in the $s$-channel. In the weak correlation regime (large $\omega$) only the highest occupied NO has a positive coefficient and all the other NOs have a negative coefficient. In the strongly correlated regime (small $\omega$) the signs of the NO are alternating. In the color version the sign of the corresponding coefficient has been coded as black for positive and red for negative.}
  \label{fig:sHarmonium}
\end{figure}

This result is orthogonal to the claim made by Cioslowski and Pernal~\cite{CioslowskiPernal2000b}. They have studied the behavior of the wavefunction coefficients of Hooke's atom numerically and found that the coefficients become very small and that the sign pattern of the most significant coefficients around these points change. Therefore, they concluded that the expansion coefficients have to become zero to change their sign. As a courtesy to the reader, we have repeated their calculations to obtain an accurate expression for the wavefunction and calculated the NO coefficients by diagonalizing the wavefunction directly. The results for the occupation numbers in the $s$-channel are shown in Fig.~\ref{fig:sHarmonium} and are identical to the ones reported in Ref.~\cite{CioslowskiPernal2000b}. Indeed, the most significant NOs have a different sign in the large and small $\omega$ limit. However, upon closer inspection of the plot in Fig.~\ref{fig:sHarmonium}, one readily sees that coefficients of \emph{different} NOs actually gain in amplitude when making the transition between the weakly and strongly correlated regime, so there is actually no evidence that the NO coefficients do cross zero. The numerical results of Cioslowski and Pernal are therefore in agreement with our findings.

%%%%%%%%%%%%%%%%%%%%%%%%%%%%%%%%%%%%%%%%
\section{Conclusion}
%%%%%%%%%%%%%%%%%%%%%%%%%%%%%%%%%%%%%%%%
The question of the existence of NOs with vanishing occupation numbers in many-body Coulomb systems is important for a number of practical applications. For the direct expansion of wavefunctions in finite orbital basis sets (CI expansions) it would be beneficial if only a finite number of NOs has a finite occupancy. However, the presence of unoccupied NOs will cause complications in the formal developments of 1RDM functional theory. For the extended Koopmans' theorem the existence of vanishing natural occupation numbers could even be catastrophic, since the ionization energies do not necessarily converge to the exact ones when the approximate wavefunction converges to the exact many-body state.

The divergence of the Coulomb interaction between the electrons requires the wavefunction to have a cusp at the coalescence points of the electrons~\eqref{eq:cuspCondition}. This non-analytic behavior can only be represented by including an infinite amount of orbitals (NOs) in the expansion of the wavefunction, so there are an infinite amount of NOs with a non-zero occupation number. However, this argument does not provide a proof that natural occupation numbers equal to zero do not exist, i.e.\ that the NOs with a non-zero occupation numbers form a complete set. However, we have been able to show that wavefunctions of the form $\alpha(\vecr_1)\alpha(\vecr_2)f(\vecr_1 - \vecr_2)$ do not have any vanishing occupation number, if and only if the Fourier transform of the correlation function, $f(\vecr_{12})$, does not vanish on an open set. The Fourier transform of the correlation function only seems to disappear if the wavefunction is separable, i.e.\ representable in a finite basis. Applying our theorem to the harmonium atom, we have shown that the occupation numbers do not vanish in contradiction with earlier assertions by Cioslowski and Pernal based on numerical calculations~\cite{CioslowskiPernal2000b}. However, a more careful inspection of their results showed that only their interpretation was incorrect and that their results actually agree with our proof up to numerical accuracy.

Further, we have demonstrated that a discontinuity in the wavefunction is not required for the absence of unoccupied NOs. Even the perfectly smooth ground state of the double harmonium has no vanishing occupation numbers. More essential is the non-separability of the wavefunction, which is caused by the discontinuity of the cusp. The discontinuity of cusp does have an effect on the decay-rate of the occupation numbers. It causes the occupation numbers to decay merely polynomially compared to an exponential decay of the occupation numbers for wavefunctions without a cusp.

\begin{acknowledgments}
The authors acknowledge the Academy of Finland for research funding under Grant No.\ 127739. KJHG also gratefully acknowledges a VENI grant by the Netherlands Foundation for Research NWO (722.012.013).
\end{acknowledgments}

\appendix

%%%%%%%%%%%%%%%%%%%%%%%%%%%%%%%%%%%%%%%%%%%%%%%%
\section{A lower bound for the decay of eigenvalues}
\label{ap:WeylTheorem}
%%%%%%%%%%%%%%%%%%%%%%%%%%%%%%%%%%%%%%%%%%%%%%%%
Though the proof of~\eqref{eq:WeylDecay} is ``\foreignlanguage{german}{äußerst einfach}'' according to Weyl~\cite{Weyl1911}, it might be worthwhile to expose its derivation. Weyl starts with an alternative derivation of a theorem by E. Schmidt~\cite{Schmidt1907} to prove that
\begin{align}\label{eq:SchmidtInequality}
\iinteg{\vecs}{\vect}\abs{K(\vecs,\vect) - k_n(\vecs,\vect)}^2 \geq \lambda_{n+1}^2 + \lambda_{n+2}^2 + \dotsb,
\end{align}
where $\lambda_k$ are the eigenvalues of the integral kernel $K(\vecs,\vect)$ ordered in descending order and $k_n(\vecs,\vect)$ is a hermitian finite rank operator
\begin{align*}
k_n(\vecs,\vect) = \sum_{i,j=1}^nk_{ij}g_i(\vecs)g_j(\vect),
\end{align*}
with $g_i \in L^2$ and $k_{ij} = k_{ji}^*$. This theorem is easily understood by using the spectral representation of the integral kernel. The minimum value of the integral is achieved by using the largest eigenvalues of $K(\vecs,\vect)$ at the diagonal, $k_{ij} = \lambda_i\delta_{ij}$ and the corresponding eigenfunctions for the functions $g_p$. This choice exactly eliminates the largest eigenvalues of $K(\vecs,\vect)$ and only the smaller $n+1$ eigenvalues will contribute to the integral. Any other choice for $k_n(\vecs,\vect)$ will give a larger value of the integral.

For more rigor, consider the following proof. To cover the infinite dimensional case we use the Rayleigh quotient to define the eigenvalues. The first eigenvalue (and largest in magnitude) is defined as
\begin{align*}
\lambda_1 \isDefinedAs \max_{f \neq 0}\frac{\norm{\hat{K}f}}{\norm{f}} = \max_{\norm{f}=1}\norm{\hat{K}f},
\end{align*}
where $\norm{\cdot}$ is the usual $L^2$ norm and the operator $\hat{K}$ is defined be the action of the integral kernel $K(\vecs,\vect)$ on a function $f$ as
\begin{align*}
\hat{K}f(\vecs) \isDefinedAs \integ{\vect}K(\vecs,\vect)f(\vect)
\end{align*}
The function that achieves this maximum, $\phi_1(\vecs)$, is the corresponding eigenfunction. The other eigenvalues are defined (found) by searching over a subspace where the previously found eigenfunctions have been projected out
\begin{align*}
\lambda_{n+1} \isDefinedAs \min_{\phi_1,\dotsc,\phi_n}\;\max_{\substack{f \perp \phi_1,\dotsc,\phi_n \\ \norm{f} = 1}}\norm{\hat{K}f}
\end{align*}
and the function that achieves this maximum, $\phi_{n+1}$, is the corresponding eigenfunction. Using this definition we readily find for the eigenvalue of the the sum of two linear operator $K_1$ and $K_2$~\cite{Rasmussen2001}
\begin{align*}
\lambda_{n+m+1}(K_1 + K_2) &= \min_{\mathclap{\phi_1,\dotsc,\phi_{n+m}}}\qquad\quad\max_{\mathclap{\substack{f \perp \phi_1,\dotsc,\phi_{n+m} \\ \norm{f} = 1}}}\,\norm{(\hat{K}_1+\hat{K_2})f} \\
&\leq \min_{\phi_1,\dotsc,\phi_{n+m}}\;\max_{\substack{f \perp \phi_1,\dotsc,\phi_n \\ \norm{f} = 1}}\norm{\hat{K}_1f} \\*
&\eqspace
{} + \min_{\mathclap{\phi_1,\dotsc,\phi_{n+m}}}\quad\;
\max_{\substack{f \perp \phi_{n+1},\dotsc,\phi_{n+m} \\ \norm{f} = 1}}\norm{\hat{K_2}f} \\
&= \min_{\phi_1,\dotsc,\phi_n}\;\max_{\substack{f \perp \phi_1,\dotsc,\phi_n \\ \norm{f} = 1}}\norm{\hat{K}_1f} \\*
&\eqspace
{} + \min_{\mathclap{\phi_{n+1},\dotsc,\phi_{n+m}}}\quad\;\;\;
\max_{\substack{f \perp \phi_{n+1},\dotsc,\phi_{n+m} \\ \norm{f} = 1}}\norm{\hat{K_2}f} \\
&= \lambda_{n+1}(K_1) + \lambda_{m+1}(K_2).
\end{align*}
Since the rank of $k_n$ is only $n$, it has only $n$ non-zero eigenvalues at maximum. Therefore, we find that
\begin{align*}
\lambda_{n+m+1}(K) \leq \lambda_{m+1}(K - k_n).
\end{align*}
Applying this inequality for all eigenvalues of $K$, we readily recover Schmidt's inequality~\eqref{eq:SchmidtInequality}.

Schmidt's inequality gives a lower bound for the integral. An upper bound can be obtained from Taylor's theorem. This procedure is probably most clearly explained at the end of Ref.~\cite{Chang1952}. To simplify the analysis, we assume without loss of generality that we integrate over a finite block with sides of length $L$, so that we can divide it in $m^d$ smaller blocks, where $d$ is the dimension of our integration variable. In each of these regions we can make a Taylor expansion around its centre $\vecs_0$
\begin{align*}
K(\vecs,\vect) = \sum_{\mathclap{\abs{\mat{p}} \leq p}}\frac{(\vecs-\vecs_0)^{\mat{p}}}{\mat{p}!}\du_{\vecs}^{\mat{p}}K(\vecs_0,\vect)  + \sum_{\mathclap{\abs{\mat{p}} = p}}h_{\mat{p}}(\vecs)(\vecs - \vecs_0)^{\mat{p}},
\end{align*}
where $\mat{p}$ denotes a multi-index
\begin{align*}
\mat{p}! &\isDefinedAs p_1! \dotsb p_d! &
\vecx^{\mat{p}} &\isDefinedAs x_1^{p_1} \dotsb x_d^{p_d} \\
\abs{\mat{p}} &\isDefinedAs p_1 + \dotsb + p_d &
\du_{\vecs}^{\mat{p}}f &\isDefinedAs \frac{\du^{\abs{\mat{p}}}f}{\du s_1^{p_1}\dotsb\du s_d^{p_d}}
\end{align*}
and the remainder satisfies
\begin{align*}
\lim_{\vecs \to \vecs_0}h_{\mat{p}}(\vecs) = 0.
\end{align*}
By choosing all possible powers of $\vecs$ up to order $p$ as basis functions for $k_n$, we can create $N = \binom{p+d}{d}$ linearly independent functions per block, so $N m^d$ functions $g_i$ in total. Using these basis functions, we can set the kernel $k_n$ with $n = Nm^d$ equal to the Taylor expansions of $K$ in these blocks. The error of the Taylor expansion can now be approximated as
\begin{align*}
\babs{K(\vecs,\vect) - k_{Nm^d}(\vecs,\vect)} \leq \epsilon_m\bigl(L/m\bigr)^p,
\end{align*}
where $\epsilon_m \to 0$ as $m \to \infty$. Now combining this inequality from the Taylor expansion of the kernel with Schmidt's inequality~\eqref{eq:SchmidtInequality} we have
\begin{multline*}
L^d\epsilon_m^2\bigl(L/m\bigr)^{2p} \geq \sum_{k=1}^{\infty}\lambda^2_{Nm^d+k} \\
\geq \sum_{k=1}^{N m^d}\lambda^2_{Nm^d+k} 
\geq m^d\lambda^2_{2Nm^d},
\end{multline*}
so for $m\to\infty$ we need
\begin{align*}
\lim_{m\to\infty}m^{2p+d}\lambda_{2Nm^d}^2 = 0.
\end{align*}
Now setting $k = 2Nm^d$ and cleaning up the limit, Weyl's inequality for the asymptotic behavior of the eigenvalues~\eqref{eq:WeylDecay} readily follows. Tighter bounds on the asymptotic decay of the eigenvalues can be found by using additional properties of the integral kernel~\cite{HilleTamarkin1931, Rasmussen2001}.

%%%%%%%%%%%%%%%%%%%%%%%%%%%%%%%%%%%%%%%%%%%%%%%%
\section{NOs for the 1D model atom}
\label{ap:modelAtom}
%%%%%%%%%%%%%%%%%%%%%%%%%%%%%%%%%%%%%%%%%%%%%%%%
First we need to determine the boundary condition to be imposed on the solutions, which will give the required quantization of the expansion coefficients $c_k$. To find the boundary conditions, we first rewrite the integral equation as
\begin{multline*}
c_k\varphi_k(x) = K\binteg{y}{-\infty}{x}\alpha^2(y)\bigl(1+\eta(x-y)\bigr)\varphi_k(y) \\
{} + K\binteg{y}{x}{\infty}\alpha^2(y)\bigl(1-\eta(x-y)\bigr)\varphi_k(y).
\end{multline*}
Now considering the limit $x \to \infty$, the last integral vanishes and we find that the solutions have to behave asymptotically as
\begin{multline*}
c_k \varphi_k(x) \sim (1 + \eta x)K\binteg{y}{-\infty}{\infty}\alpha^2(y)\varphi_k(y) \\
{} - \eta K\binteg{y}{-\infty}{\infty}\alpha^2(y) y \varphi_k(y),
\end{multline*}
since the orbital should decay exponentially for the wavefunction to be normalizable. Because we are dealing with a symmetric orbital, the solutions can be separated in gerade and ungerade functions. For the even solutions only the first contribution survives, so we have the following boundary condition for $x \to \infty$ for the gerade solutions
\begin{align}\label{eq:uEvenLargeX}
\varphi_{g,k}(x \to \infty) 	&\sim \frac{1 + \eta x}{c_k}K\binteg{y}{-\infty}{\infty}\alpha^2(y)\,\varphi_{g,k}(y).
\end{align}
Likewise, for the ungerade solutions only the last term survives and we find
\begin{align}\label{eq:uOddLargeX}
\varphi_{u,k}(x \to \infty) 	&\sim \frac{\eta K}{c_k}\binteg{y}{-\infty}{\infty}\alpha^2(y)\,y\,\varphi_{u,k}(y).
\end{align}
Note that this analysis of the boundary conditions is completely general for symmetric $\alpha^2(x)$. Now we have found the boundary conditions for the various solutions, we turn our attention to the solution of the differential equation
\begin{align*}
\varphi''(x) = \lambda \e^{-2Z\abs{x}}\varphi(x),
\end{align*}
where the normalization constant is calculated to be
\begin{align*}
\frac{\eta K}{Z^2} = \frac{2}{\sqrt{4\,Z^2/\eta^2+6\,Z/\eta + 4}}.
\end{align*}
Since we are looking for symmetry adapted solutions, we only need to solve this differential equation for $x \geq 0$. If we define (for $x \geq 0$)
\begin{align}\label{eq:sDef}
s(x) = \frac{\sqrt{\abs{\lambda}}}{Z}\e^{-Z x},
\end{align}
then for a function $\varphi(x) = f\bigl(s(x)\bigr)$ we have
\begin{align*}
\frac{\ud \varphi}{\ud x} &= \frac{\ud f}{\ud s}\frac{\ud s}{\ud x} = -\sqrt{\abs{\lambda}}\e^{-Zx}f'(s) = -Zs f'(s) \\
\frac{\ud^2\varphi}{\ud x^2} &= f''(s)\left(\frac{\ud s}{\ud x}\right)^{\mathrlap{2}} + f'(s)\frac{\ud^2 s}{\ud x^2} \\
&= Z^2\bigl(s^2f''(s) + sf'(s)\bigr)
\end{align*}
and the differential equation in term of $f(s)$ becomes
\begin{align*}
s^2f''(s) + sf'(s) - \sgn(\lambda)s^2f(s) = 0.
\end{align*}
This is the Bessel differential equation for the zeroth order Bessel functions and the general solution to this equation is
\begin{align*}
f(s) &= C_1\BesselJ_0(s) + C_2\BesselY_0(s)	&	&\text{for $\lambda < 0$}, \\
f(s) &= C_1\BesselI_0(s) + C_2\BesselK_0(s)	&	&\text{for $\lambda > 0$},
\end{align*}
with $C_1$ and $C_2$ constants and $\BesselJ_0$ and $\BesselY_0$ Bessel functions of the first and second kind respectively and $\BesselI_0$ and $\BesselK_0$ their modified counterparts. Now going back to the original function $u(x)$ we find for $\lambda < 0$
\begin{align*}
\varphi^-(x) 
= C_1\BesselJ_0\left(\frac{\sqrt{-\lambda}}{Z}\e^{-Zx}\right) + C_2\BesselY_0\left(\frac{\sqrt{-\lambda}}{Z}\e^{-Zx}\right)
\end{align*}
and in the case of $\lambda > 0$ we find
\begin{align*}
\varphi^+(x) 
= C_1\BesselI_0\left(\frac{\sqrt{\lambda}}{Z}\e^{-Zx}\right) + C_2\BesselK_0\left(\frac{\sqrt{\lambda}}{Z}\e^{-Zx}\right).
\end{align*}
Now we construct the even and odd solutions by imposing the corresponding boundary conditions. First we impose the boundary conditions at $x=0$. The odd solutions need to vanish at the origin, so we find
\begin{align*}
\varphi^-_{u}(x) &= C\bigl[\BesselY_0\bigl(\tilde{\lambda}\bigr)\BesselJ_0\bigl(s(x)\bigr) -
\BesselJ_0\bigl(\tilde{\lambda}\bigr)\BesselY_0\bigl(s(x)\bigr)\bigr], \\
\varphi^+_{u}(x) &= C\bigl[\BesselK_0\bigl(\tilde{\lambda}\bigr)\BesselI_0\bigl(s(x)\bigr) -
\BesselI_0\bigl(\tilde{\lambda}\bigr)\BesselK_0\bigl(s(x)\bigr)\bigr],
\end{align*}
where $\tilde{\lambda} \coloneqq \sqrt{\abs{\lambda}}/Z$ and $C$ is a normalization constant. For the even solutions the first order derivative at $x=0$ needs to vanish, so for the even solutions we find
\begin{align*}
\varphi^-_{g}(x) &= C\bigl[\BesselY_1\bigl(\tilde{\lambda}\bigr)\BesselJ_0\bigl(s(x)\bigr) -
\BesselJ_1\bigl(\tilde{\lambda}\bigr)\BesselY_0\bigl(s(x)\bigr)\bigr], \\
\varphi^+_{g}(x) &= C\bigl[\BesselK_1\bigl(\tilde{\lambda}\bigr)\BesselI_0\bigl(s(x)\bigr) +
\BesselI_1\bigl(\tilde{\lambda}\bigr)\BesselK_0\bigl(s(x)\bigr)\bigr].
\end{align*}
To obtain the proper quantization of the eigenvalue $\lambda$, we need to impose the proper boundary conditions for $x \to \infty$, i.e.\ $s \to 0$. From the asymptotic behavior of the Bessel functions for small $s$, we find that our ungerade solutions behave asymptotically for $x \to \infty$ as
\begin{align*}
\varphi^-_{u}(x) &\sim C\left[\BesselY_0\bigl(\tilde{\lambda}\bigr) - \frac{2}{\pi}\BesselJ_0\bigl(\tilde{\lambda}\bigr)
\Bigl(\gamma + \ln\bigl(\tilde{\lambda}/2\bigr) - Zx\Bigr)\right], \\
\varphi^+_{u}(x) &\sim C\left[\BesselK_0\bigl(\tilde{\lambda}\bigr) + \BesselI_0\bigl(\tilde{\lambda}\bigr)
\Bigl(\gamma + \ln\bigl(\tilde{\lambda}/2\bigr) - Zx\Bigr)\right],
\end{align*}
where $\gamma$ is the Euler--Mascheroni constant. We found before, however, that the odd solutions do not have a linear term~\eqref{eq:uOddLargeX}, so we must have
\begin{align*}
\BesselJ_0\bigl(\sqrt{-\lambda}/Z\bigr)	&= 0	&&\text{for $\lambda < 0$}, \\
\BesselI_0\bigl(\sqrt{\lambda}/Z\bigr) 	&= 0	&&\text{for $\lambda > 0$}.
\end{align*}
Since $\BesselI_0(y)$ does not have any zero, only solutions for $\lambda < 0$ exist, which are related to the zeros of the the zeroth order Bessel function of the first kind, $\BesselJ_0(y_k) = 0$, as
\begin{align*}
\lambda_{u,k} = -\bigl(Z y_k\bigr)^2.
\end{align*}
Using~\eqref{eq:lambdaDef} the coefficients of the odd NOs are readily determined to be
\begin{align*}
c_{u,k} = -\frac{\eta K}{Z^2}\frac{2}{y_k^2}.
\end{align*}
The zero's of the zeroth order Bessel function behave asymptotically as
\begin{align*}
y_k = \pi\Bigl(k - \frac{1}{4}\Bigr) \qquad (k \to \infty),
\end{align*}
so asymptotically, the coefficients decay quadratically, $c_{u,k} = \Order\bigl(k{-2}\bigr)$. To construct the corresponding NOs, we use that $u_{u,k}(-x) = - u_{u,k}(x)$, so
\begin{align*}
\phi_{u,k}(x) = C_{u,k}\sgn(x)\e^{-Z\abs{x}}\BesselJ_0\Bigl(y_k\e^{-Z\abs{x}}\Bigr).
\end{align*}
The normalization constant of the ungerade NOs is readily calculated by using that the $n^{\text{th}}$ order Bessel functions satisfy
\begin{align*}
\binteg{\rho}{0}{1}\rho\BesselJ_n(\alpha_{nj}\rho)\BesselJ_n(\alpha_{nk}\rho) 
= \half\bigl(\BesselJ_{n+1}(\alpha_{nj})\bigr)^2\delta_{jk},
\end{align*}
where $\alpha_{nj}$ is the $j^{\text{th}}$ zero of $\BesselJ_n$. Making the substitution $\rho = \e^{-Z x}$ and taking $n = 0$, we find that the normalization constant for the odd NOs is given as
\begin{align*}
C_{u,k} = \frac{\sqrt{Z}}{\BesselJ_1(y_k)}.
\end{align*}
Now we will construct the even solutions. First we consider the large $x$ behavior of the even solutions. From the asymptotic behavior of the Bessel functions for small $s$ we find that for $x \to \infty$
\begin{align*}
\varphi^-_{g}(x) &\sim C\Bigl[\BesselY_1\bigl(\tilde{\lambda}\bigr) - \frac{2}{\pi}\BesselJ_1\bigl(\tilde{\lambda}\bigr)
\Bigl(\gamma + \ln\bigl(\tilde{\lambda}/2\bigr) - Zx\Bigr)\Bigr], \\
\varphi^+_{g}(x) &\sim C\Bigl[\BesselK_1\bigl(\tilde{\lambda}\bigr) - \BesselI_1\bigl(\tilde{\lambda}\bigr)
\Bigl(\gamma + \ln\bigl(\tilde{\lambda}/2\bigr) - Zx\Bigr)\Bigr].
\end{align*}
Since we know that the the even solutions have to behave for $x \to \infty$ as given by the asymptotic relation in~\eqref{eq:uEvenLargeX}, we must have that the the ratio between the linear and constant term must be equal to $\eta$, i.e.\ $f^{\pm}\bigl(\tilde{\lambda}\bigr) = 0$, where
\begin{align*}
f^+(y) &\isDefinedAs 1 - \frac{\eta}{Z}\left[\frac{\BesselK_1(y)}{\BesselI_1(y)} - \bigl(\gamma + \ln(y/2)\bigr)\right], \\
f^-(y) &\isDefinedAs 
\BesselJ_1(y)
 - \frac{\eta}{Z}\left[\frac{\pi}{2}\BesselY_1(y) - \BesselJ_1(y)\bigl(\gamma + \ln(y/2)\bigr)\right].
\end{align*}
Note that in $f^+$ we could divide by $\BesselI_1(y)$, since this is an exponentially growing function and has no zeros apart from $y=0$. 

\begin{figure}[t]
  \includegraphics[width=\columnwidth]{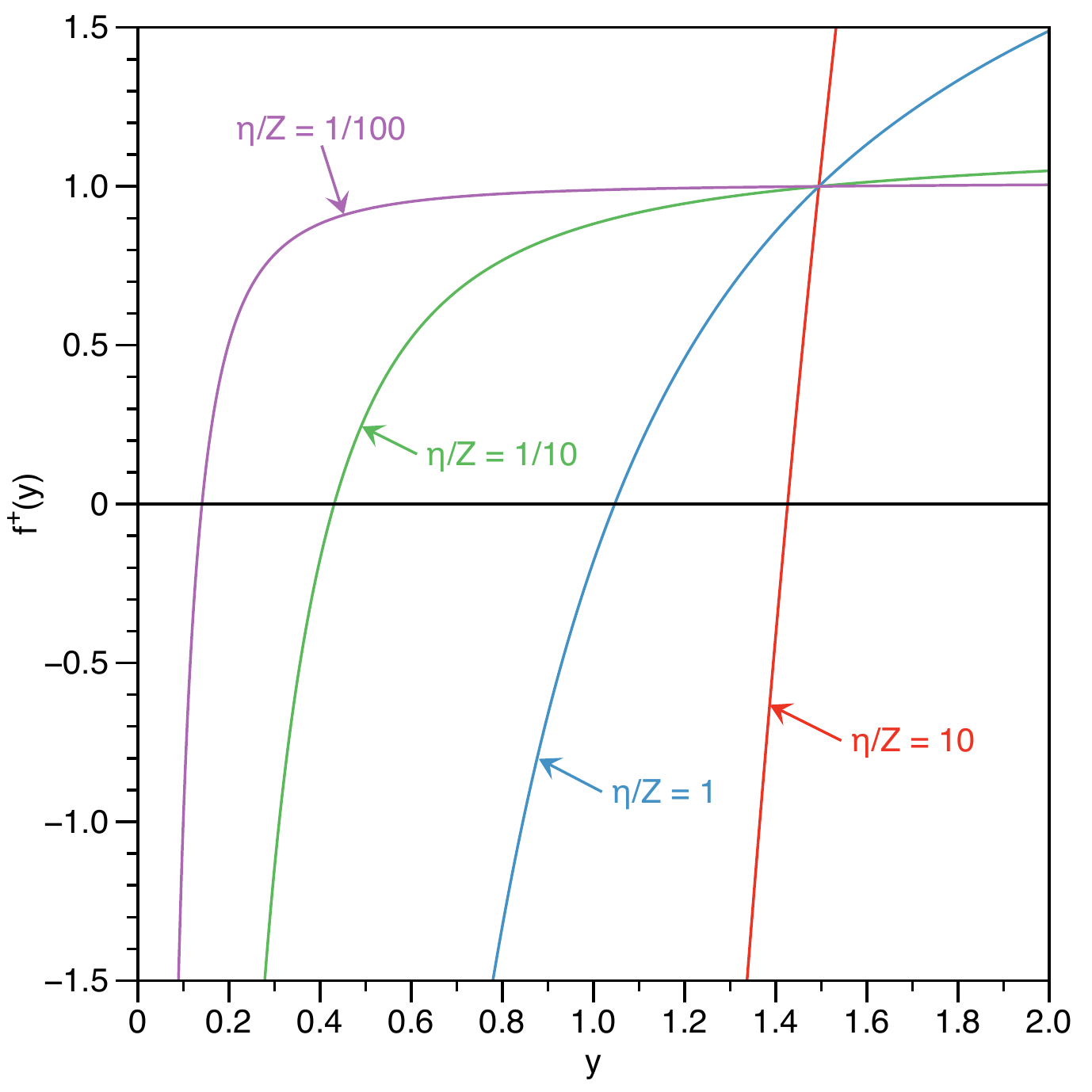}
  \caption{Plots of $f^+(y)$ for four different values of $\eta/Z = 10, 1, 1/10, 1/100$.}
  \label{fig:fPlus}
\end{figure}

Let us now first consider the positive coefficients, $\lambda > 0$. 
The function $f^+(y)$ is constructed out of modified Bessel functions of the first and second kind, which are monotonically decreasing and increasing functions respectively. Since the logarithm is a monotonically increasing function, the function $f^+(y)$ is also a monotonically increasing function, so $f^+(y)$ can have only one zero at most. Because the ratio of the modified Bessel functions $\BesselK_1(y)/\BesselI_1(y)$ diverges at $y \to 0$ as $1/y^2$, we have $f^+(y \to 0) = -\infty$. Further, due to the logarithm $f^+(y)$ diverges logarithmically for $y \to \infty$, so $f^+(y)$ has always exactly one zero (see Fig.~\ref{fig:fPlus}). Hence we find that there is exactly one NO with a positive coefficient of gerade symmetry. The corresponding NO, $\phi^+_g(x)$, is readily constructed as
\begin{multline*}
\phi^+_g(x) 
= C^+_g\biggl[\biggl(\frac{Z}{\eta} + \gamma + \ln\Bigl(\frac{y_0}{2}\Bigr)\biggr)\BesselI_0\Bigl(y_0\e^{-z\abs{x}}\Bigr)  \\*
{} + \BesselK_0\Bigl(y_0\e^{-Z\abs{x}}\Bigr)\biggr]\e^{-Z\abs{x}},
\end{multline*}
where $C^+_g$ is a normalization constant and $y_0$ is the zero of the function $f^+(y)$. The corresponding attains the simple expression
\begin{align*}
c^+_g = \frac{\eta K}{Z^2}\frac{2}{y_0^2}.
\end{align*}
The negative solutions, $\lambda < 0$, can be obtained in a similar manner. The only difference is that the function $f^-(y)$ has an infinite amount of zeros, $z_k$, due to the infinite amount of oscillations of the usual Bessel functions. Following the same steps as for the positive orbital, we find that the gerade NOs with negative coefficients are
\begin{multline*}
\phi^-_{g,k} = C^-_{g,k}\biggl[\biggl(\frac{Z}{\eta} + \gamma + \ln\Bigl(\frac{z_k}{2}\Bigr)\biggr)
\BesselJ_0\Bigl(z_k\e^{-Z\abs{x}}\Bigr) \\*
- \frac{\pi}{2}\BesselY_0\Bigl(z_k\e^{-Z\abs{x}}\Bigr)\biggr]\e^{-Z\abs{x}}
\end{multline*}
where $C^-_{g,k}$ are normalization constants and $z_k$ are the zeros of the function $f^-(y)$ defined. The corresponding (negative) coefficients are
\begin{align*}
c^-_{g,k} = -\frac{\eta K}{Z^2}\frac{2}{z_k^2}.
\end{align*}
For large $k$, $z_k$ is a large number, so asymptotically the condition $f^-(z) = 0$ becomes
\begin{align*}
\tan\Bigl(z - \frac{3\pi}{4}\Bigr)
= \frac{2}{\pi}\left(\frac{Z}{\eta} + \gamma + \ln(z/2)\right).
\end{align*}
For very large $z$ the logarithm on the right-hand side will diverge and the tangent on the left-hand side is only large when $z = (4n+1)\pi/4$ for an integer $n$. So we find that the the zeros behave asymptotically as
\begin{align*}
z_k = \pi\Bigl(k + \frac{1}{4}\Bigr) \qquad (k \to \infty),
\end{align*}
so the negative coefficients of the even NOs decay quadratically, $c^-_{g,k} = \Order\bigl(k^{-2}\bigr)$, exactly as the coefficients of the odd NOs.

%%%%%%%%%%%%%%%%%%%%%%%%%%%%%%%%%%%%%%%%%%%%%%%%
\section{NOs for harmonic potential and interaction}
\label{ap:harmNOs}
%%%%%%%%%%%%%%%%%%%%%%%%%%%%%%%%%%%%%%%%%%%%%%%%
In this appendix we will show how to find the NOs for the ground state of a system with only harmonic interactions (Sec.~\ref{sec:harmInt}). The NOs satisfy the following integral equation
\begin{align*}
\integ{y}\Psi(x,y)\phi_k(y) = c_k\phi_k(x),
\end{align*}
where the integral kernel is simply the wavefunction which is for the current purpose most conveniently written as
\begin{align*}
\Psi(x,y) = \sqrt[4]{\frac{\omega\tilde{\omega}}{\pi^2}}\e^{-\alpha(x^2+y^2) - 2\beta x y},
\end{align*}
where $\alpha \isDefinedAs (\omega + \tilde{\omega})/4$ and $\beta \isDefinedAs (\omega - \tilde{\omega})/4$. Since all the interactions are harmonic, we expect that a Gaussian, $\phi_0 = \e^{-\nu x^2}$, might be a solution. Working out the integral gives
\begin{align*}
\integ{y}\Psi(x,y)\e^{-\nu x^2} = \sqrt[4]{\frac{\omega\tilde{\omega}}{\pi^2}}
\sqrt{\frac{\pi}{\alpha+\nu}}\e^{-\left(\alpha - \frac{\beta^2}{\alpha + \nu}\right)x^2},
\end{align*}
so we find that a Gaussian is indeed an eigenfunction if the exponent is set to
\begin{align*}
\nu = \sqrt{\alpha^2 - \beta^2} = \half\sqrt{\omega\tilde{\omega}}.
\end{align*}
The corresponding eigenvalue can be worked out to be
\begin{align*}
c_0 = \frac{2\sqrt[4]{\omega\tilde{\omega}}}{\bigl(\sqrt{\omega} + \sqrt{\tilde{\omega}}\bigr)}.
\end{align*}
The other solutions will now be of the form $\phi_k(x) = P_k(x)\e^{-\half\sqrt{\omega\tilde{\omega}}x^2}$, where $P_k(x)$ is a polynomial of order $k$. From the eigenvalue equation, it follows that these polynomials satisfy
\begin{align*}
\sqrt[4]{\frac{\omega\tilde{\omega}}{\pi^2}}\integ{y}P_k(y)\e^{-\frac{1}{4}\left((\sqrt{\omega}+\sqrt{\tilde{\omega}})y + (\sqrt{\omega}-\sqrt{\tilde{\omega}})x\right)^2} = c_kP_k(x).
\end{align*}
Now we make the following coordinate transformation $u \isDefinedAs \half\bigl(\sqrt{\omega} + \sqrt{\tilde{\omega}}\bigr)x$ and $v \isDefinedAs \half\bigl(\sqrt{\omega} + \sqrt{\tilde{\omega}}\bigr)y$, which simplifies the equation for the polynomials to
\begin{align}
\integ{v}Q_k(v-\eta u)\e^{-v^2} = \mu_kQ_k(u),
\end{align}
where
\begin{align*}
\eta &\isDefinedAs \frac{\sqrt{\omega} - \sqrt{\tilde{\omega}}}{\sqrt{\omega} + \sqrt{\tilde{\omega}}}, \\
\mu_k &\isDefinedAs \half\bigl(\sqrt{\omega}+\sqrt{\tilde{\omega}}\bigr)
\sqrt[4]{\frac{\pi^2}{\omega\tilde{\omega}}}c_k, \\
Q_k(x) &\isDefinedAs P_k\biggl(\frac{2x}{\sqrt{\omega}+\sqrt{\tilde{\omega}}}\biggr).
\end{align*}
Solving the equation for the polynomials of the first four orders gives
\begin{align*}
Q_0(u) &= 1						&\mu_0 &= \sqrt{\pi}, \\
Q_1(u) &= u						&\mu_1 &= \sqrt{\pi}(-\eta), \\
Q_2(u) &=1 + 2(\eta^2-1)u^2			&\mu_2 &= \sqrt{\pi}(-\eta)^2, \\
Q_3(u) &= u + \tfrac{2}{3}(\eta^2-1)u^3	&\mu_3 &= \sqrt{\pi}(-\eta)^3.
\end{align*}
We see that the eigenvalues are simply related by a factor $-\eta$, so we should be able to recover this relation. Taking the derivative $n$ times from the integral equation for $Q_k$, we find
\begin{align*}
\mu_kk!q_k = (-\eta)^kk!q_k\integ{v}\e^{-v^2} = \sqrt{\pi}(-\eta)^kk!q_k,
\end{align*}
where $q_k$ denotes the highest order coefficients of the polynomial $Q_k$. Hence we find that the eigenvalues are simply $\mu_k = \sqrt{\pi}(-\eta)^k$, which can be worked out to give~\eqref{eq:harmCoefs}.

Due to the integration Gaussian as integration weight, we expect that the polynomials are related to the Hermite polynomials. Using the following definition for the Hermite polynomials,
\begin{align*}
\Hermite_k(x) \isDefinedAs (-1)^k\e^{x^2}\bigl(\du_x\bigr)^k\e^{-x^2},
\end{align*}
it follows from our results for the polynomials $Q_0(u)$ \ldots $Q_3(u)$ that they should be related to the Hermite polynomials as $Q_k(u) = \Hermite_k(\sqrt{1-\eta^2}u)$. Since the Hermite polynomials satisfy the recursion relation
\begin{align*}
\Hermite_{k+1}(x) = 2x\Hermite_k(x) - \Hermite'_k(x),
\end{align*}
the polynomials $Q_k(u)$ should satisfy
\begin{align*}
Q_{k+1}(u) = 2\sqrt{1-\eta^2}\,u\,Q_k(u) - \frac{Q_k'(u)}{\sqrt{1-\eta^2}}.
\end{align*}
Inserting the recursion relation in the integral equation for $Q_n(u)$, we find that $Q_{k+1}(u)$ is indeed also a solution if $Q_k(u)$ is a solution, with an eigenvalue $\mu_{k+1} = -\eta\cdot\mu_k$. Hence, we have found all NOs in terms of the Hermite polynomials. Working out them out in the original quantities, one recovers the expression for the NOs for the harmonic interaction~\eqref{eq:harmNOs}.

%%%%%%%%%%%%%%%%%%%%%%%%%%%%%%%%%%%%%%%%%%%%%%%%
\section{Fourier transforms}
\label{ap:Fouriers}
%%%%%%%%%%%%%%%%%%%%%%%%%%%%%%%%%%%%%%%%%%%%%%%%
In this appendix we show in more detail how the Fourier transforms of the correlation functions of the various model systems were calculated. Let us first consider the correlation functions of the 1D systems. In the case of the double-harmonium the correlation function is a Gaussian, so the Fourier transform should be again a Gaussian. Indeed, when we work it out, we find
\begin{align*}
\Fourier\Bigl[\e^{-\half(\tilde{\omega} - \omega)x^2}\Bigr](k)
&= \binteg{x}{-\infty}{\infty}\e^{\I k x}\e^{-\half(\tilde{\omega} - \omega)x^2} \\
&= \e^{-\half\frac{k^2}{\tilde{\omega}-\omega}}\binteg{x}{-\infty}{\infty}
\e^{-\half(\tilde{\omega} - \omega)\left(x - \frac{\I k}{\tilde{\omega}-\omega}\right)^2} \\
&= \e^{-\half\frac{k^2}{\tilde{\omega}-\omega}}\binteg{y}{-\infty}{\infty}\e^{-\half(\tilde{\omega} - \omega)y^2} \\
&= \sqrt{\frac{2\pi}{\tilde{\omega}-\omega}}\e^{-\half\frac{k^2}{\tilde{\omega}-\omega}},
\end{align*}
where we used that the integration contour could be shifted through the complex plane, since there are no poles.

For the Fourier transform of the 1D Hylleraas wavefunction we have
\begin{multline*}
\Fourier[1+\eta\abs{x}](k)
= \binteg{x}{-\infty}{\infty}\e^{\I k x}\bigl(1+\eta\abs{x}\bigr) \\
= 2\pi\delta(x) + \eta\lim_{\alpha \to 0^+}\binteg{x}{0}{\infty}x\bigl(\e^{\I k x} + \e^{-\I k x}\bigr)\e^{-\alpha x},
\end{multline*}
where we introduced the $\alpha$-limit to make the integral convergent. This integral is readily worked out by differentiating under the integral sign
\begin{align*}
\binteg{x}{0}{\infty}x&\bigl(\e^{\I k x} + \e^{-\I k x}\bigr)\e^{-\alpha x} \\
&= -\du_{\alpha}\binteg{x}{0}{\infty}\bigl(\e^{(\I k -\alpha)x} + \e^{-(\I k -\alpha)x}\bigr) \\
&= -\du_{\alpha}\frac{2\alpha}{\alpha^2 + k^2}
= 2\frac{\alpha^2 - k^2}{(\alpha^2 + k^2)^2}.
\end{align*}
Now taking the limit $\alpha \to 0$, we find for the full Fourier transform as in~\eqref{eq:Hylleraas1Dfourier}.

The other systems are three dimensional and have in common that their correlation function only depends on the length, $r_{12}$, so we can already do the integration over the angles
\begin{align*}
\Fourier[f(r)](k)
&= \integ{\vecr}\e^{\I \veck\cdot\vecr}f(r) \\
&= 2\pi\binteg{r}{0}{\infty}r^2f(r)\binteg{s}{-1}{1}\e^{\I k r s} \\
&= \frac{4\pi}{k}\binteg{r}{0}{\infty}r \sin(kr)\, f(r).
\end{align*}
Let us first consider the Fourier transform of the correlation function of the 3D Hylleraas wavefunction
\begin{align*}
\Fourier[1+\eta r](k) &= \frac{4\pi}{k}\binteg{r}{0}{\infty}r \sin(kr)\,\bigl(1 + \eta r\bigr).
\end{align*}
For the first integral we have
\begin{align}\label{eq:constFourier3D}
\binteg{r}{0}{\infty}r \sin(kr)
&= -\du_k\binteg{r}{0}{\infty}\cos(kr) \notag \\
&= -\frac{\du_k}{2}\binteg{r}{-\infty}{\infty}\e^{\I k r}
= -\pi\du_k\delta(k).
\end{align}
For the second integral we need to introduce the convergence factor again, so we evaluate
\begin{align*}
\binteg{r}{0}{\infty}r^2\sin(kr)\e^{-\alpha r}
&= \frac{\du_{\alpha}^2}{2\I}\binteg{r}{0}{\infty}\bigl(\e^{(\I k - \alpha)r} - \e^{-(\I k + \alpha)r}\bigr) \\
&= \du_{\alpha}^2\frac{k}{\alpha^2 + k^2}
= 2k\frac{3\alpha^2 - k^2}{(\alpha^2 + k^2)^3}
\end{align*}
Using this result in the limit $\alpha \to 0$ together with~\eqref{eq:constFourier3D}, we find the full Fourier transform of the 3D Hylleraas correlation as in~\eqref{eq:Hylleraas3Dfourier}.

Finally let us consider the Fourier transform for the correlation function of the system with inverse harmonic interactions. For general $\alpha > 0$ we have
\begin{align*}
\Fourier[r^{\alpha}](k)
&= \frac{4\pi}{k}\lim_{\eta\to0^+}\binteg{r}{0}{\infty}r^{1 + \alpha} \sin(kr)\e^{-\eta r} \\
&= \frac{2\pi}{\I k}\lim_{\eta\to0^+}\binteg{r}{0}{\infty}r^{1 + \alpha} \bigl(\e^{-(\eta - \I k)r} - \e^{-(\eta + \I k) r}\bigr) \\
&= \frac{2\pi}{\I k}\lim_{\eta\to0^+}
\Biggl(\frac{1}{(\eta - \I k)^{2 + \alpha}}\binteg{t}{0}{(\eta - \I k)\infty}\quad t^{\alpha + 1}\e^{-t} - {} \\*
&\eqspace\hphantom{\frac{2\pi}{\I k}\lim_{\eta\to0^+}\Biggl(}
\frac{1}{(\eta + \I k)^{2 + \alpha}}\binteg{t}{0}{(\eta + \I k)\infty}\quad t^{\alpha + 1}\e^{-t}\Biggr) \\
&= \frac{2\pi}{\I k^{3+\alpha}}\left(\e^{(2+\alpha)\pi\I/2} - \e^{-(2 + \alpha)\pi\I/2}\right)\binteg{t}{0}{\infty} t^{\alpha + 1}\e^{-t} \\
&= \frac{4\pi}{k^{3+\alpha}}\sin\left(\frac{2+\alpha}{2}\pi\right)\Gamma(2+\alpha) \\
&= -\frac{4\pi}{k^{3+\alpha}}\sin(\pi\alpha/2)\Gamma(2+\alpha),
\end{align*}
where we used that the integration interval could be deformed without difficulty, since the integrant does not have any poles, and that the definition of the gamma function
\begin{align*}
\Gamma(z) \coloneqq \binteg{t}{0}{\infty}t^{z - 1}\e^{-t}.
\end{align*}
We find that the Fourier transform diverges for $k \to 0$. In the case of even $\alpha$, however, it is not clear what happens, since we obtain a division of zero by zero. To asses what the real answer should be, we have to calculate these cases separately. For even $\alpha$, we can work out the Fourier transform by taking successive derivatives under the integral
\begin{align*}
\Fourier[r^{\alpha}](k)
&= \frac{4\pi}{k}\binteg{r}{0}{\infty}r^{1 + \alpha} \sin(kr) \\
&= \frac{4\pi}{k}\bigl(-\du^2_k\bigr)^{\alpha/2}\binteg{r}{0}{\infty}r \sin(kr) \\
&= \frac{4\pi^2}{k}(-1)^{\alpha/2}\delta^{(1+\alpha)}(k)
\end{align*}
where $\delta^{(n)}(k)$ denotes the $n^{\text{th}}$ order derivative of the delta-function and we used the previous result for the 3D Fourier transform of a constant function~\eqref{eq:constFourier3D}. Combining these results, we find for general $\alpha > 0$ the Fourier transform as reported~\eqref{eq:rAlphaFourier}. Note that this result for general $\Fourier[r^{\alpha}](k)$ can also be used to obtain the Fourier transform of the correlation function of the 3D Hylleraas wavefunction and give the same result.

%%%%%%%%%%%%%%%%%%%%%%%%%%%%%%%%%%%%%%%%%%%%%%%%
\section{Laplace transform of the correlation function of the Hookium atom}
\label{ap:hookLaplace}
%%%%%%%%%%%%%%%%%%%%%%%%%%%%%%%%%%%%%%%%%%%%%%%%
Here we will consider a series solution for the differential equation of the Laplace transform of the correlation function of the Hookium atom~\eqref{eq:hookLaplEq}. To construct the series, first observe that a physical solution requires, $\lim_{s \to \infty} \tilde{t}(s) = 0$, otherwise the inverse Laplace transform does not exist~\cite{ArfkenWeberHarris2013}. Hence, the solution should be expanded in powers of $1/s$ rather than in powers of $s$. Using the Frobenius trick, we find that the solution for $\tilde{t}(s)$ can be expanded as
\begin{align*}
\tilde{t}(s) = \frac{1}{s^2}\sum_{\nu=0}^{\infty}\frac{\tilde{a}_{\nu}}{s^{\nu}},
\end{align*}
where the coefficients $\tilde{a}_{\nu}$ satisfy the following recursion relation
\begin{align*}
\tilde{a}_0 &\neq 0, \\
\tilde{a}_1 &= \frac{\lambda}{\sqrt{\omega/2}}\tilde{a}_0, \\
\tilde{a}_{\nu} &= \frac{\lambda}{\sqrt{\omega/2}}\frac{\tilde{a}_{\nu-1}}{\nu} + 
\left(2\nu - 1 - \frac{2\epsilon_t}{\omega}\right)\tilde{a}_{\nu - 2}
\end{align*}
The second boundary condition is effectively given by the condition that the full wavefunction is normalizable~\cite{Taut1993}. This additional condition serves as a quantization condition for the internal energy contribution, $\epsilon_r$. In general, the power series will not terminate. However, there are special ratios $\lambda^2/\omega$ at which the series terminates after a finite amount of terms. Since the series truncates after a finite amount of terms at these ratios, the wavefunction is automatically normalizable~\cite{Taut1993}.

The series solution for $\tilde{t}(s)$ could also have been obtained by Laplace transforming the series solution for $t(\rho)$ by Taut~\cite{Taut1993} term-by-term using that
\begin{align*}
\tilde{p}_m(s) = \Laplace[r^{m+1}](s) = \binteg{r}{0}{\infty}r^{m+1} = \frac{(m+1)!}{s^{m+2}}.
\end{align*}
Similar to the series solution of $t(\rho)$~\cite{Taut1993}, the series solution of $\tilde{t}(s)$ has at least two terms due to the cusp condition. As a check consider the solution for $\lambda^2/\omega = 2$, which has only two terms
\begin{align*}
\tilde{t}(s) = \frac{1}{s^2} + \frac{1}{s^3}.
\end{align*}
Using~\eqref{eq:FourLapl}, we can construct the Fourier transform of the correlation function, $f$. From the first term we get up to the factor $2\pi\I/k$
\begin{align*}
\lim_{\sigma\to 0^+}\biggl(&\frac{1}{(\sigma+\I k)^2} - \frac{1}{(\sigma-\I k)^2}\biggr) \\
&= \lim_{\sigma\to 0^+}\frac{-4\I\sigma k}{(\sigma^2 + k^2)^2}
= 2\I\lim_{\sigma\to 0^+}\du_k\frac{\sigma}{(\sigma^2 + k^2)} \\
&= 2\I\du_k\lim_{\sigma\to 0^+}\frac{\sigma}{(\sigma^2 + k^2)}
= 2\pi\I\delta'(k),
\end{align*}
where we used that the delta function
\begin{align*}
\delta(x) = \frac{1}{\pi}\lim_{\eta \to 0}\frac{\eta}{x^2 + \eta^2}.
\end{align*}
The second term is easier, since we can drop the limit immediately to give
\begin{align*}
\frac{2\pi\I}{k}\left(\frac{1}{(\I k)^3} - \frac{1}{(-\I k)^3}\right)
= -\frac{4\pi}{k^4}.
\end{align*}
Combing these results, we find
\begin{align*}
-\frac{4\pi}{k}\left(\frac{1}{k^3} + \pi\delta'(k)\right) = \Fourier\bigl[1 + \thalf r\bigr](k),
\end{align*}
where the identification with the Fourier transform of $1 + \half r$ was readily made by comparing with the Fourier transform of the correlation function of the 3D Hylleraas wavefunction~\eqref{eq:Hylleraas3Dfourier}.

\end{document}